\documentclass[12pt]{article}
\usepackage{amssymb,amsmath,framed}
\usepackage[dvips]{graphicx}

\catcode `\@=11
\def\@cite#1#2{#1\if@tempswa , #2\fi}
\catcode `\@=12

\newcommand{\qed}{\hbox{\rule[-2pt]{3pt}{6pt}}}

\begin{document} 

\title{An equivalence relation of boundary/initial conditions, \\
       and the infinite limit properties.} 
\author{{\sc Kazuhiko Minami} \\ \\
  {\small{\it Graduate School of Mathematics, Nagoya University,}}\\
  {\small{\it  Nagoya, 464-8602, JAPAN}}
}

\maketitle  

The 'n-equivalences' of boundary conditions of lattice models 
are introduced  
and it is derived that the models with n-equivalent boundary conditions 
result in the identical free energy.  
It is shown 
that the free energy of the six-vertex model 
is classified through the density of left/down arrows on the boundary. 
The free energy becomes identical to that 
obtained by Lieb and Sutherland with the periodic boundary condition, 
if the density of the arrows is equal to $1/2$. 
The relation to the structure of the transfer matrix 
and a relation to stochastic processes are noted. 
\\

\vspace{1.2cm}


minami@math.nagoya-u.ac.jp
\\
\newpage


Two configurations $\{ s_i\}$ and $\{ s'_i\}$ 
where the corresponding local variables are the same: $s_i=s'_i$, 
except a finite number of them 
can be regarded as being identical when we consider the thermodynamic limit properties. 
This is also true when the differences approach infinity 
but remain of the lower orders than bulk quantities 
such as the total number of sites or the total energy contributions. 
It follows that boundary conditions, for example, of lattice models 
are usually irrelevant to the thermodynamics. 

On the other hand, a six-vertex model with a specific boundary condition was solved [\cite{1}] 
and it is found that the free energy of it is not identical to 
the free energy previously obtained [\cite{2,3,4}] with the cyclic
boundary condition.  This difference of two free energies comes from the
fact  that the configurations of the six-vertex model should satisfy the
six-vertex restriction, and therefore
the boundary condition is still relevant in the thermodynamic limit, 
even if the energy contributions coming from the boundary sites 
are irrelevant in the limit. 

In this short note 
an equivalence relation of boundary conditions is introduced.  
It is generally derived that 
two boundary conditions yield the same free energy 
if they belong to the same equivalence class, 
even in models with long-ranged restrictions.  
Several related properties will also be discussed.


First let us define the equivalence relation. 
Let us consider classical lattice models 
where each local variables takes one of a finite number of states. 
We assume that the number of boundary sites 
is of a lower order than the total number of sites. 
Let us consider the set of sites 
which lie in the lattice and can be reached from the boundary sites 
by just $n$ steps ($n$ bonds) at minimum; 
we call these sites the $n$-boundary sites. 
Let us consider the set of bonds 
between $(n-1)$- and $n$-boundary sites, 
and call them the $n$-boundary bonds. 
At last let us consider the $n$-boundary sites together with the $n$-boundary bonds 
and call them the $n$-boundary, 
and also call the configurations on the $n$-boundary the $n$-boundary configurations. 
Let $\{ \Gamma_i\}$ be the set of all the possible configurations on the $n$-boundary 
with a boundary condition $\Gamma$ on the actual boundary of the lattice. 
Two boundary conditions $\Gamma$ and $\Gamma'$ 
are called $n$-equivalent when $\{ \Gamma_i\}=\{ \Gamma'_i\}$ 
as a set of $n$-boundary configurations. 


Then we can show the following fact. 
Suppose that the boundary conditions $\Gamma$ and $\Gamma'$ 
are $n$-equivalent throughout the thermodynamic limit with some finite $n$. 
Then the two free energies with $\Gamma$ and $\Gamma'$ are identical. 
Proof:
Let us write the partition function as $Z=\sum_i B_iZ_i$. 
The factor $Z_i$ is the partition function from the set of local variables 
on the lattice inside the $n$-boundary with fixed $n$-boundary configuration $\Gamma_i$. 
The factor $B_i$ is the partition function from the other local variables 
with $\Gamma$ and $\Gamma_i$. 
Then the factor $B_iZ_i$ is the partition function of the system 
with the $\Gamma$ and fixed $\Gamma_i$. 
From the assumption that $n$ is always finite, 
we obtain  
$\log Z_i =-\beta Nf_i+o(N)$ and $\log B_i =o(N)$ 
where $N$ is the total number of sites. 
The index $i$ runs from $1$ to $i_{\rm max}$ 
where $i_{\rm max}$ is the number of the possible configurations 
on the $n$-boundary with the boundary condition $\Gamma$. 
We have $i_{\rm max}\leq O(r^{N'})\;$ 
where $r$ is a constant and $N'$ is the number of sites on the $n$-boundary.  
Then we obtain
\begin{eqnarray}
\frac{1}{N}\log Z&=&\frac{1}{N}\log(\sum_iB_iZ_i)
\nonumber\\
&=&\frac{1}{N}
\log B_1Z_1[1+\sum_{i=2}^{m_1}\frac{B_i}{B_1}\frac{Z_i}{Z_1}
             +\sum_{i=m_1+1}^{i_{\rm max}} e^{-\beta N(f_i-f_1)+o(N)}]
\nonumber\\
&\rightarrow& -\beta f_1  \hspace{0.6cm} (N\rightarrow\infty),
\nonumber
\end{eqnarray} 
where $f_i=f_1 \; (1\leq i\leq m_1)$, $f_1 < f_i \; (m_1+1 \leq i)$.
\nonumber
\qed


All the boundary conditions are 1-equivalent to each other for the Ising models  
because the spin states are independent of their nearest-neighbors. 
This results in the following trivial fact that the free energies of the Ising models 
with finite range interactions are independent of the boundary conditons.


This equivalence can be used in a slightly generalized form. 
We can concentrate on a part of the boundary 
({\it e.g.} one of the four edges of the rectangle, this will be considered below) 
with a condition $\Gamma$, 
and introduce the corresponding $n$-boundary and the corresponding $n$-equivalence.


Let us consider the six-vertex model on a rectangular lattice 
with $h$ rows and $w$ columns. 
We assign an arrow on each bond. 
Two arrows in and the other two out at each site. 
Then there exist six types of possible local arrow arrangements as shown in Fig.1, 
and we call them the six vertices. 
Each vertex has finite energy, the energy is assumed to be unchanged 
reversing all the four arrows around the site. 
One can assign a line to a bond if the arrow on the bond points down or points left, 
as shown again in Fig.1, 
and there exists a one-to-one correspondence between configurations of arrows 
and configurations of continuous lines on the lattice. 
The possible six arrow arrangements satisfy the restriction 
that the lines do not intersect each other 
and continue beginning from one bond on the boundary of the lattice 
to another bond on the boundary. 
The total number of lines is hence conserved 
when the boundary configuration is fixed.

Let us assume 
that the line configuration on the first row of vertical bonds 
(upper edge of the rectangle) 
is fixed and identical to that on the last row, 
and also assume 
that the line configuration on the first column of horizontal bonds 
(the right edge of the rectangle) 
is fixed and identical to that on the last column. 
In this case the number of lines 
is conserved in each step moving from the first row to the last row, 
and also from the first column to the last column. 
Let us assume that there are always 
$n_1$ ($n_2$) lines on the first $p_0$ boundary bonds on the upper (right) edge 
and $n_1$ ($n_2$) lines on the next $p_0$ bonds on the edges, 
and vice versa. 
In this case we introduce the boundary line density $\rho_i$ 
as $\rho_i=n_i/p_0$ where $i=1$ and $2$. 
Fix a sequence of finite rectangles approaching to the thermodynamic limit. 
Then the free energy of the six-vertex model 
is determined by $\rho_1$ and $\rho_2$: $f=f(\rho_1,\rho_2)$. 
Proof: 
If $\rho_2=0$ the free energy is equal to $(1-\rho_1)\epsilon_1+\rho_1\epsilon_2$ 
(and similarly if $\rho_1=0$). 
As for the cases with the densities equal to one, 
we can use the fact that 
the free energy with $(\rho_1,\rho_2)=(1-\rho,1-\rho')$ 
is identical to that with $(\rho_1,\rho_2)=(\rho,\rho')$, 
which comes from the symmetry of the vertex energies. 
If both of $\rho_1$ and $\rho_2$ satisfy $0<\rho_i<1$, 
consider an edge with the density $\rho_1$ and the corresponding $n$-boundary. 
It is easy to convince that 
all the boundary line configurations on the edge with the density $\rho_1$ 
are $n$-equivalent to each other, 
where $n=kp_0$ is now a fixed constant proportional to $p_0$. 
This argument is valid respectively for the other three edges. 
\qed

It can be derived [\cite{5}], with the use of the $n$-equivalence, 
in the case of the six-vertex model with more general situation, 
that $f(1/2,1/2)=f_{\rm LS}$  
where $f_{\rm LS}$ is the free energy obtained [\cite{2,3,4}] 
with the cyclic boundary condition 
in both the horizontal and the vertical directions.


This argument can be applied to line conserved models on other lattices. 
The argument is also easily generalized to models 
with $l_i$ lines on each bond $i$, where $l_i=0, 1, \ldots, q-1$, 
that is the $q$ state vertex models. 
On the square lattice, for example, with $q=3$ we have the 19-vertex model. 

A kind of SOS model is equivalent [\cite{6}] to a $q$ state vertex model 
and also equivalent to a class of spin $S$ classical and quantum spin chains, 
where the total magnetization corresponds to the number of lines.


Next we note the relation 
between the $n$-equivalence and the structure of the transfer matrix. 
Let us consider the six-vertex model on the rectangle 
and assume the cyclic boundary condition in the horizontal direction, 
i.e. the six-vertex model on a cylinder.  
The free energy is obtained 
from the maximum eigenvalue of the transfer matrix $V$. 
All the matrix elements of $V$ are non-negative 
because each element is a sum of Boltzmann weights. 
The matrix $V$ is block-diagonalized according to the number of lines $m$ 
which is conserved from row to the next row, 
and let $V_m$ be the corresponding block element. 
It is easy to convince that 
all the line configurations with $m$ lines, $m\leq w$, on the upper edge 
are $n$-equivalent to each other, 
where $n=kw$ and $k$ is a positive integer. 
This means that 
there are allowed line configurations on the lattice 
with arbitrary fixed boundary line configurations with $m$ lines on the first row 
and other arbitrary line configurations with $m$ lines on the last row, 
if $h$ is larger than $2kw$. 
Then all the matrix elements of $V_m^{h}$ are positive, 
and hence $V_m^{h}$ is irreducible, 
because any matrix is irreducible if all of its elements are positive. 
We can thus conclude that $V_m$ is irreducible, 
because $V_m^{h}$ cannot be irreducible if $V_m$ is not. 
This fact means that the transfer matrix is irreducible 
if the bases (line configurations on the row) 
used for the matrix representation 
are $n$-equivalent to each other. 
We can also say 
that all the line configurations on the first row with $m$ lines 
result in the same free energy 
in the limit $h\to\infty$ with fixed $w$, 
because they are $kw$-equivalent and $kw$ is finite. 
The free energies remains of course identical 
when we take the limit $w\to\infty$ afterwards.


At last we will make a comment on the relation with stochastic processes. 
In this case, $V$ is a stochastic matrix 
where the sum of the elements of each row is unity 
that is the total sum of probabilities. 
It is known that 
the maximum eigenvalues of such matrix are always unity 
and the absolute values of the other eigenvalues do not exceed unity. 
Let us consider 
a stochastic system characterized by a finite number of variables $(l_1, \ldots, l_s)$, 
with a kind of conservation law $\sum l_i=$const. 
This kind of stochastic processes 
correspond to the line-conserved lattice models. 
The stochastic matrix is block diagonalized for each set of $n$-equivalent initial conditions. 
Each block element is irreducible 
and the limiting state is again classified according to the $n$-equivalence.


\newpage 

\begin{center}
\includegraphics[width=12cm,clip]{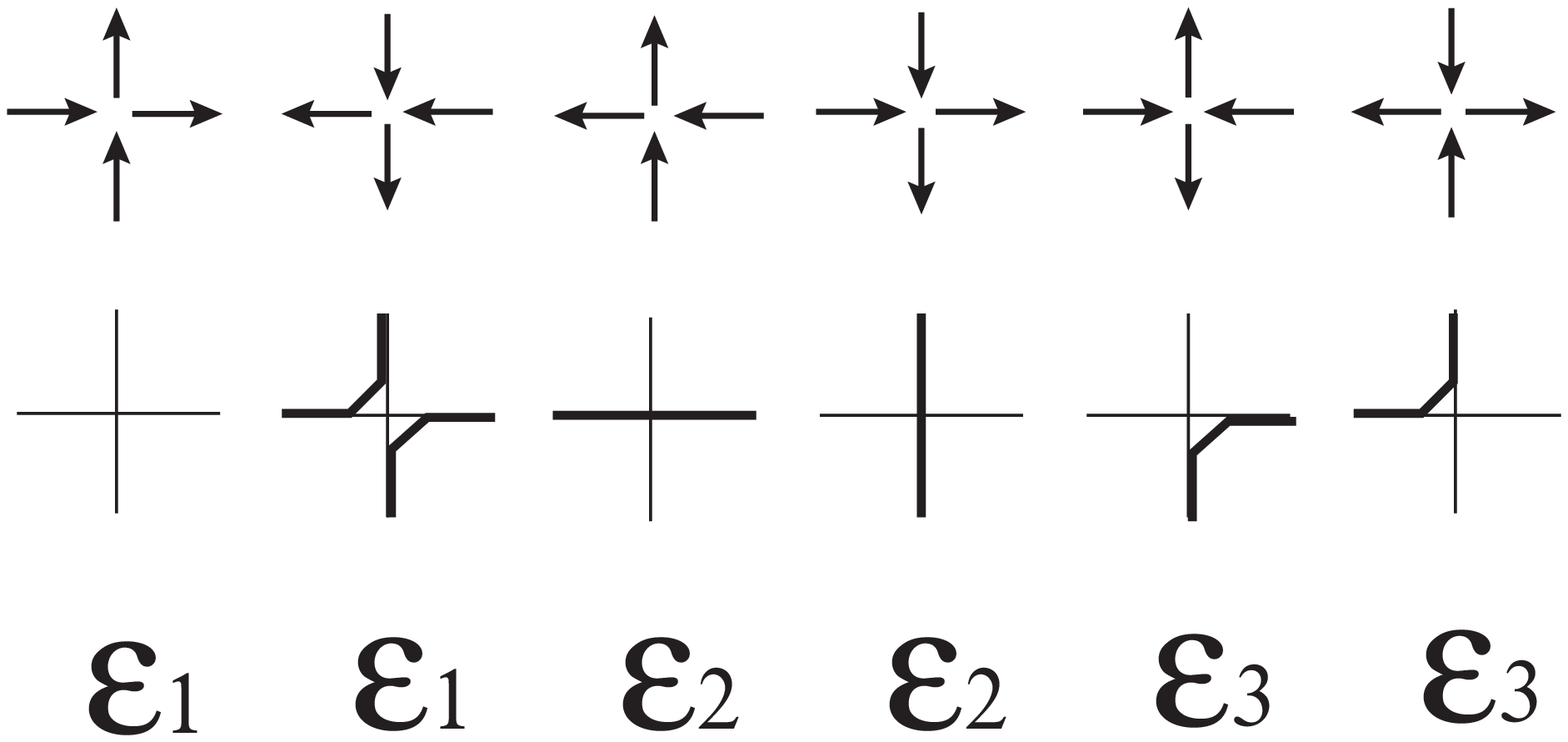} \\
Fig.1\:  Six vertices, corresponding lines and energies.
\end{center}

\end{document}